\documentclass[aps,prb,twocolumn,superscriptaddress,showpacs,floatfix]{revtex4-1}
\usepackage{graphicx}
\usepackage{amsmath}
\usepackage{amssymb}
\usepackage{url}
\usepackage{float}
\usepackage{color}
\usepackage{natbib}

\usepackage{bm}

\graphicspath{{.}{../}{./eps/}}

\newcommand{\ket}[1]{\left| #1 \right>} 
\newcommand{\bra}[1]{\left< #1 \right|} 
\newcommand{\braket}[2]{\left< #1 \vphantom{#2} \right|
 \left. #2 \vphantom{#1} \right>} 
\let\baraccent=\= 
\renewcommand{\=}[1]{\stackrel{#1}{=}} 

\newcommand{\medio}[1]{\left\langle #1 \right\rangle}

\def\XXint#1#2#3{{\setbox0=\hbox{$#1{#2#3}{\int}$}
     \vcenter{\hbox{$#2#3$}}\kern-.5\wd0}}

\newcommand{\p}[1]{^{#1}}

\renewcommand{\exp}[1]{\text{e}^{#1}}

\begin{document}

\title{ Adatoms and Anderson Localization in Graphene}

\author{Jose H. Garc\'ia}
\affiliation{Instituto de F\'isica, Universidade Federal do Rio de Janeiro, Caixa Postal 68528,  Rio de Janeiro, RJ 21941-972, Brazil}
\author{Bruno Uchoa}
\affiliation{Department of Physics and Astronomy, University of Oklahoma, 440 W. Brooks Street, Norman, Oklahoma 73019, USA}
\author{Lucian Covaci}
\affiliation{Department Fysica, Universiteit Antwerpen, Groenenborgerlaan 171, B-2020 Antwerpen, Belgium}
\author{Tatiana G. Rappoport}
\affiliation{Instituto de F\'isica, Universidade Federal do Rio de Janeiro, Caixa Postal 68528, 21941-972 Rio de Janeiro RJ, Brazil}

\date{\today}

\begin{abstract}
We address the nature of the disordered state that results from the adsorption of adatoms in graphene.  For adatoms that sit at the center of the honeycomb plaquette, as in the case of most transition metals, we show that the ones that form a zero-energy resonant state lead to Anderson localization in the vicinity of the Dirac point.  Among those, we show that there is a symmetry class of adatoms where Anderson localization is suppressed,  leading to an exotic metallic state with large and \emph{rare} charge droplets, that localizes \emph{only} at the Dirac point.  We identify the experimental conditions for the observation of the Anderson transition for adatoms in graphene. 
\end{abstract}


\maketitle

\section{Introduction}
The diffusive motion of electrons in metals can be strongly affected by disorder. For instance, disorder can localize electrons and produce an insulating state, a phenomenon known as Anderson localization (AL) ~\cite{anderson}. For noninteracting electrons,  the one parameter scaling theory predicts that electrons must localize for any arbitrary strength of short-range disorder in two spatial dimensions \cite{abrahams_scaling_1979}. Graphene \cite{graphene}, as other known  low-dimensional systems \cite{Philip}, has an unconventional localization phenomenology.  At weak coupling, depending on the type of disorder, preserved  symmetries such as chirality and the absence of backscattering between valleys can prevent localization and lead the system to a quantum critical metal-insulator transition~\cite{mirlinPRB, caio}. In experiment, the evidence of AL in graphene remains elusive \cite{Geim}. Predictions based on lattice models indicate that AL is possible in the presence of vacancies \cite{vitor} and  strong scalar disorder \cite{vitor, KT}. Top carbon site resonant scatterers, which preserve chirality \cite{mirlin2010},  and weak Coulomb impurities do not lead to localization \cite{aires, mirlin2010, wehling}. 

In this paper, we describe the problem of localization for a disordered distribution of adatoms sitting at the center of the honeycomb hexagons ($H$ site), as in the case of most transition metals \cite{chan_first-principles_2008}. In this configuration, the adatoms mediate hopping processes between distant carbon sites in the plaquette of the impurity \cite{bruno2011b, weeks11} and explicitly break the chiral symmetry of the Hamiltonian at the lattice scale. We propose an effective graphene-only Hamiltonian for disordered graphene and conduct a scaling analysis of the local density of states (LDOS) for large system sizes. We show that the formation of zero-energy resonant states in the plaquette of the impurity  leads to AL in the vicinity of the Dirac point and to a metal-insulator transition at a well defined energy, which defines the  mobility edge. We find that AL appears in two distinct classes, depending on the symmetry of the resonant orbitals. In particular, when each of the orbitals that form the resonant state preserves the sublattice  point group symmetry of graphene, destructive  interference between the different hybridization paths in the plaquette of the adatom leads to a different anomalous class of localization. In this class,  the system shows an exotic nonhomogeneous metallic state with large and rare  charge droplets, that localizes \emph{only} at the Dirac point.  We propose that the Anderson transition  can be observed and characterized with scanning tunneling spectroscopy (STS) probes \cite{STMscience}. We indicate the most promising adatoms that can produce AL in graphene.  

In Sec. \ref{sec-si} we consider the effective graphene only Hamiltonian, which we derive in the Appendixes.  In Sec. III we establish the conditions for the appearance of zero-energy resonant states in graphene with adatoms sitting at $H$ sites.  In Se. IV, we present a numerical method to characterize electronic localization and analyze the problem of AL for a disordered distribution of adatoms with different possible orbital symmetries.

 \section{\label{sec-si} Effective Hamiltonian}
To capture the physics of localization, we start from  the electronic Hamiltonian of graphene in the presence of a single adatom, which can be described by the non-interacting Anderson Hamiltonian,  $\mathcal{H}=\mathcal{H}_g+ \mathcal{H}_f+ \mathcal{H}_V$, where 
\begin{align}
\mathcal{H}_g=-t\!\sum_{\medio{i,j},\sigma} c^\dagger_{ \sigma} (\mathbf{R}_i) c_{ \sigma} (\mathbf{R}_j)  
\end{align}
is the graphene Hamiltonian.  $t\approx$ 2.8eV is the hopping energy between nearest neighbor (NN) sites in a honeycomb lattice, and $c_{\sigma}(\mathbf{R}_i)$  is  the annihilation operator of electrons with spin $\sigma=\uparrow,\downarrow$ on  site $\mathbf{R}_i$. The second term in $\mathcal{H}$ represents the Hamiltonian of the localized electrons at 
the impurity site, $\mathcal{H}_f= \epsilon_0\sum_{m,\sigma} f^\dagger_{m\sigma}f_{m\sigma}$, where $\epsilon_0$ is the energy of the localized state and $  f_{m \sigma}$ is the annihilation operator for the localized electrons with spin $\sigma$ in a given irreducible representation with angular momentum $l$ and angular momentum projection $m\leq l$. The sum over $m$ is carried over \emph{all} degenerate orbitals. The third term describes the hybridization term  \cite{bruno2011, bruno2011b}
\begin{align}
\mathcal{H}_{\text{V}}=\sum_{\sigma,m}\sum_{i \in I} V_{i}^{(m)}c_{\sigma}^{\dagger}(\mathbf{R}_{i})f_{m,\sigma}(\mathbf{R}_{I})+H.c.
\end{align}
 where $V_{i}$ are the hybridization amplitudes of the adatom with each of the NN carbon atoms in the honeycomb plaquette $I$, centered at the coordinate $\mathbf{R}_{I}$, as shown in Fig. 1a.   Almost all heavy atoms are likely to hybridize at the $H$ site, and most of them hybridize with graphene via $s$, $d$ and $f$ orbitals \cite{chan_first-principles_2008}. For  $m=0$ states, such as $s$ and $d_{z^2}$ orbitals,   the adatom hybridizes equally with all  the six neighboring  carbon atoms,  $V_{i \in I}^{(0)}=V$. For in-plane $f$ wave orbitals, such as $f_{x(x^2-3y^2)}$ ($|m|=3$), there is a $\pi$ phase difference in the hybridization of the adatom with the two different sublattices,   $V_{1,3,5}=-V_{2,4,6}=V$ (see Fig.1). For  a $d_{xy}$ orbital ($m=2$), $V_{1,4}=0$ and $V_{2,5}=-V_{3,6}=\sqrt{3}V/2$, while for a $d_{x^2-y^2}$ orbital ($m=-2$), $V_{1,4}=V$, and  $V_{i}=-V/2$, for $i\neq1,4$.    The hybridization amplitudes are dictated by symmetry only \cite{bruno2011b}.  

Integrating out the localized fermions, the effective Hamiltonian of graphene is the presence of a random distribution of $N$ adatoms is given by 
\begin{equation}
\mathcal{H}_{\mathrm{eff}}=\mathcal{H}_g +\sum_{I=1}^N \mathcal{H}_I, \label{Heff} 
\end{equation}
where the second term describes the effective  plaquete potential of the impurities, 
\begin{align}
\mathcal{H}_I=\!\!\!\! \sum_{ (i,j)\in I, \sigma }\!\!\!\!\tau_{ij} \, c_{\sigma}\p{\dagger}(\mathbf{R}_i)c_{\sigma}(\mathbf{R}_j)  +\delta \mu\!\sum_{ i \in I }\ \!\!\hat{n} (\mathbf{R}_i), 
\label{HI}
\end{align}
as shown in the Appendix B. The bracket  $(i,j)\in I$ in the first term indicates that the sum has to be performed over the six carbon atoms surrounding a given $H$-site impurity.  In leading order in the hybridization, the  hopping processes mediated by the impurity have the effective  amplitude 
\begin{equation}
\tau_{ij}=-(1/\epsilon_0) \sum _m {V_{i}\p{(m)}{V_{j}^{*}}\p{(m)}},
\end{equation}
including diagonal processes where the electron hops into the impurity and then back to the same site ($i=j$).   $\delta\mu$ accounts for a local charge transfer between the adatom and the six carbon atoms in the plaquette of the impurity, with $\hat{n}$ a density operator. The effective Hamiltonian hence describes graphene in the presence of a  special kind of random scalar potential, combined with  hopping processes mediated by the impurity that connect all six vertices of the honeycomb plaquette around the impurity to each other, as shown in Fig. 1a. This  ``plaquette disorder" potential allows for equally probable hopping between NN, next-nearest neighbors (NNN) and next-next-nearest neighbors (NNNN) sites, depending on the symmetry of the localized orbital. NNN hopping terms explicitly break the chiral symmetry of the Hamiltonian, permitting the emergence of AL at the Dirac point.     

Because hybridization is mediated by hopping into a virtual site of the lattice ($H$-site), the electrons acquire a phase as they hop in and out of the impurity. In $C_{3v}$ invariant orbitals, those phases destructively interfere \cite{bruno2011b} and the impurity tends to decouple from the bath, making zero-energy resonant levels (midgap states)  in that class ineffective as a source of AL for states \emph{away} from the Dirac point.    
This class of resonant orbitals, described by  $m=0$  ($s$, $d_{z^2}$, etc)  and in-plane  $f$-wave orbitals, corresponds to a plaquette where all hopping processes have the same amplitude, up to a sign (symmetric plaquette). The asymmetric  class (plaquette)  is described by  $|m|=1,\,2$  $d$-wave and $f$-wave orbitals, and the corresponding degenerate doublet states.

\section{Resonant condition}

In order to calculate the electronic properties of the Hamiltonian of eq. \ref{Heff} , we use the kernel polynomial method \cite{weise_kernel_2006}. In this method we rescale the Hamiltonian to $\tilde{H}$ so that $\tilde{E}_k\in(-1,1)$ is  the rescaled energy for all $k$ labeling a state of the Hamiltonian.  We then represent  the required spectral function as a finite series  of Chebyshev polynomials $T_m(\tilde{E})\equiv  \cos\left[m \arccos (\tilde{E})\right]$ where its expansion coefficients are calculated with sparse matrix vector multiplications.

 \begin{figure}
     \centering\includegraphics[width=1.0\columnwidth,clip]{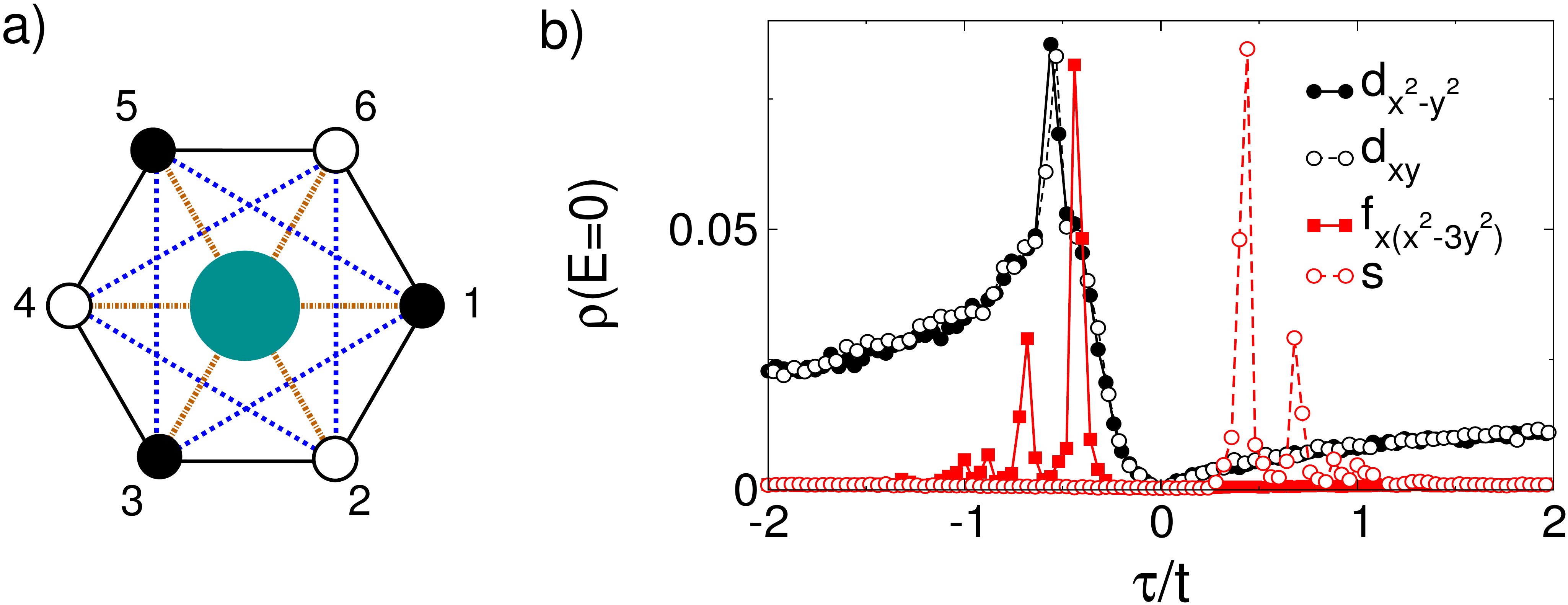}
     \caption{ a) Impurity plaquette for an adatom (center) sitting at an $H$ site, with six carbon atoms: white circles (sublattice $A$); black circles (sublattice $B$). Hopping processes mediated by the adatom: Solid lines (NN hopping), dashed (NNN hopping) and dot-dashed (NNNN hopping) (see text).   b): DOS at the Dirac point vs.  effective hopping parameter $\tau=-V^2/\epsilon_0$ for different orbital symmetries at  $p=0.05$ adatoms per carbon and  $D=2\times 1200 \times1200$ sites. 
     }
     \label{fig:simetrias}
\end{figure}

By using the definition of the LDOS 
\begin{align}
\rho_i=\frac{1}{\pi}\sum_{k}\left|\braket{k}{i}\right|^2 
\end{align}
we may expand it as follows:
\begin{align}
\rho_i=\frac{1}{\sqrt{1-\tilde{E}^2}}\left[\mu_0g_0+2\sum_{m=1}^{\infty}\mu_m g_mT_m(\tilde{E}) \delta(E-\tilde{E}_k)\right]
\end{align}
where $\mu_m=\bra{i}T_m(\tilde{H})\ket{i}$ and $g_m$ is  the Jackson Kernel, which acts as a regularization factor  that accounts for the Gibbs oscillations \cite{Dunham_approximation_1912}. 

For the numerical calculations,  we used video cards using \begin{footnotesize}CUDA-CUSP\end{footnotesize} libraries with double precision. Our systems have up to $D=2.4\times10^6$ sites and the polynomial expansion uses up to $M=3000$ moments, scanning 500 sites in 250 realizations of disorder,  in a total of $10^5$ sites for the statistics.

In Fig.  \ref{fig:simetrias}b, we plot the density of states (DOS) at the Dirac point $\rho(E=0)$ as a function of the effective hopping $\tau\equiv - V^2/\epsilon_0$  for different orbital symmetries.  In all cases, the behavior of $\rho(0)$ with $\tau$ is non-monotonic and shows a peak at $|\tau| \sim 0.5 t$, which describes the condition for resonant scattering at the Dirac point. 

This  condition can be derived through the single impurity Anderson problem, as shown in Appendix A, and corresponds to the pole of the Green's function of the localized electrons at zero energy, 
$
G_{f}^{-1}(0)=  -\epsilon_0 -\Sigma_f (0)=0, 
$
 where $\Sigma_f(E)\propto V^2$ is the self-energy due to the conduction electrons.  The resonance condition is given by
 \begin{equation}
 \tau_0=V^2 /\mathrm{Re}\Sigma_f(0),
\end{equation} 
with $\tau_0=\pm0.425t$ for $s$-wave and in-plane $f$-wave orbitals respectively, and  $\tau_0=-0.56t$ for  $d_{xy}$ and $d_{x^2-y^2}$ orbitals, degenerate or not. For top carbon adatoms, $\mathrm{Re}\Sigma_f(0)=0$ at the Dirac point, and hence the resonant condition is $\epsilon_0=0$, as in the case of vacancies. 

 The width of the peaks in Fig. \ref{fig:simetrias}b is set by the level broadening  $\Delta(0) = \mathrm{Im}\Sigma_f(0)$, which is finite in the asymmetric class, because of the enhanced DOS at the Dirac point due to the disorder, and is \emph{exactly} zero for symmetric orbitals, due to destructive interference among hybridization paths in the impurity plaquette.  Near the Dirac point, the width of the resonance for a single impurity scales with energy as $\Delta(E)\propto \pi V^{2}E^{\eta}\rho(E)$, with $\eta=0$  in the asymmetric plaquete class and $\eta=2$ in the symmetric one \cite{bruno2011b}. Although  symmetric adatoms hybridize more weakly with the electronic bath at finite energy due to interference effects, their scattering resonance, on the contrary,  becomes singularly \emph{strong} at zero energy. In this symmetry class, the Anderson transition - addressed in the next section -  is quantum critical as a function of energy at the Dirac point.

\section{Localization \label{sec-na}}

In this section, we first present in detail a method to determinate the localization properties of an electronic system though local quantities, such as the LDOS. We then apply this method to analyze the electronic properties of graphene decorated with a random distribution of adatoms sitting at $H$-sites. 

\subsection{Numerical analysis}

We analyze the probability distribution function of the local density of states $f[\rho_i(E)]$, for a fixed energy  to address the problem of Anderson transition \cite{mirlin_distribution_1994, schubert_distribution_2010,schubert_anderson_2009}. This function contains information  about  the probability to obtain a LDOS in a range $(\rho_i,\rho_i+d\rho_i)$ and can be captured within a histogram of  LDOS for a fixed energy on a random sample of sites of the lattice. When most of the states are extended, they are spread on the whole lattice and  $\rho_i$ is almost constant for all sites with a small variation due to disorder. As a result,  the local distribution function $f(\rho_i)$ is size-independent, self-averaging and gaussian-like. In this case,  the mean or average density of states for a system of $D$ atoms [which coincides with density of states $\rho(E)$],  
\begin{align}
\rho (E)=\frac{1}{D}\sum_{n=1}^{D} \rho_i(E),
\end{align}
 is the same as the geometric or typical density of states $\rho_{\text {typ}} $, defined as
\begin{align}
\rho_{\text{typ}}(E)=\text{exp}\left[\frac{1}{D} \sum_{n=1}^{D}\ln \rho_i(E)\right].
\end{align}
Hence, for extended states,  $\rho_{\text{typ}}(E) \approx \rho(E)$ and any of them is a good indicator of the global behavior of the system \cite{schubert_distribution_2010, zhang_geometric_2011}. When most states are localized, the density of states is highly concentrated in few lattice sites.  In this case, the local density of states $\rho_i$ is a strongly fluctuating quantity and its distribution is size-dependent  and non self-averaging. Rare events, characterized by large localization peaks, dominate $\rho$ and consequently shift its value in relation to $\rho_{\text{typ}}$. 

The ratio $R(E)$ between the typical and mean density of states:
\begin{align}
R(E)=\frac{\rho_{\text{typ}}(E)}{\rho(E)}\in[0,1] .
\end{align}
is  intrinsically related to $f[\rho_i(E)]$ and  has a very characteristic behavior whether the states are localized or extended.  $R(E)\sim 1$ for metallic states while $R(E)<1$ for localized ones. Furthermore,  in the latter,  $R(E)\to0$ in the thermodynamic limit, for increasing system sizes, whereas for metallic states, $R(E)$ is size independent.  In the case of AL, it is known from non-linear sigma models 
that $f(\rho)$ matches a log-normal distribution~\cite{mirlin_statistics_2000,schubert_distribution_2010}.  
In  two dimensions, the tail of the distribution scales with the system size according to $f[\rho_i(E)]\propto \mathrm{exp}(\ln^2\rho/\ln L )$~\cite{mirlin_statistics_2000}.

\subsection{Plaquette impurity potential}
In the following, we analyze the plaquette impurity potential (\ref{HI}) keeping the impurity concentration fixed at $p=0.05$ impurities per site. 
For  moderate disorder, $\delta\mu \lesssim t$  at $p=0.05$ adatoms per carbon, $\delta\mu$ renormalizes the energy of the localized state  $\epsilon_0$~\cite{pereira2}, and can be absorbed into the definition of  the first term in the plaquette potential (\ref{HI}) with renormalized hopping amplitudes, $\tau_{ij}$. In this regime, $\delta \mu$ is an irrelevant operator in the renormalization group sense and can be set to zero.

 \begin{figure}
      \centering\includegraphics[width=1\linewidth,clip]{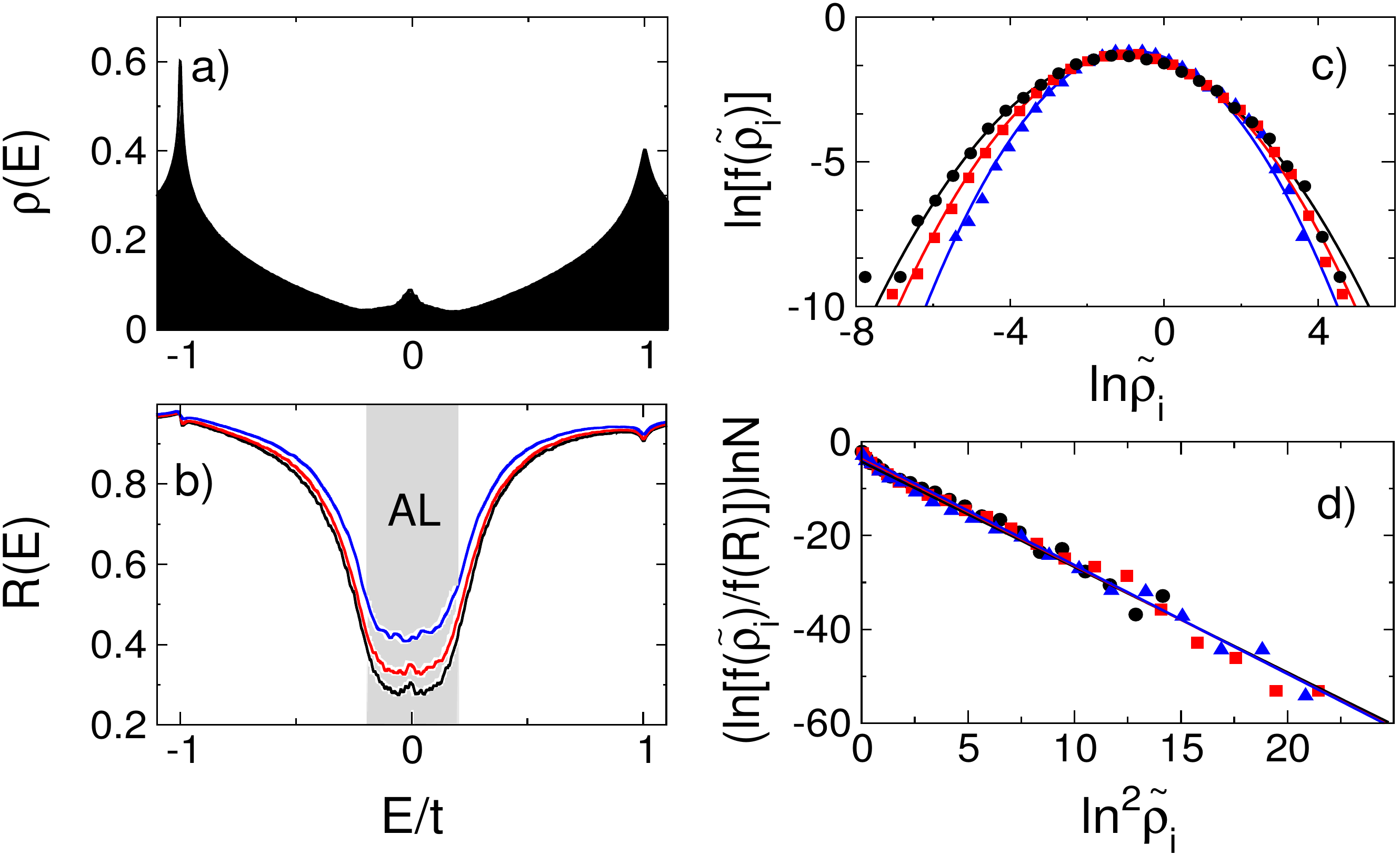}
     \caption{ (a) DOS  in the asymmetric plaquette case, for $d$-wave resonant orbitals. (b) $R(E)$ for different system sizes.   AL states are indicated in the gray region.   (c) Log of distribution function of  normalized LDOS for $E=0.1t$. (d) Finite size scaling of the tail of the distribution functions.  The sizes in panels b-d are $D=2\times L\times L$ with $L=300$ (blue), 600 (red) and 1200 (black). }
     \label{fig:dxydos}
\end{figure}

To analyze the asymmetric case, we consider a distribution of adatoms with non-degenerate and randomly oriented $d$-wave orbitals ($|m|=2$).  In Fig.  \ref{fig:dxydos}a, we plot  the density of states as a function of  energy near the resonant condition $\tau\sim-0.56t$, and a clear peak appears at the Dirac point.  To identify the nature of the peak, we calculated $R(E)$  for three different system sizes (see Fig. \ref{fig:dxydos}b).  $R(E)<1$ on the whole energy range and its minimum  is a plateau that coincides with the energy range where  the peak  in the density of states emerges (orange arrow). The width of the peak is $\sim 2 v /\ell$, where $v\sim 6$eV$\mathrm{\AA}$ is the Fermi velocity and $\ell \propto 1/\sqrt{p}$ is the average distance among the impurities, which scales with the impurity concentration $p$ \cite{vitor}.   In the energy range of the plateau, indicated in the gray region in Fig.  \ref{fig:dxydos}b, $R(E)$ is strongly reduced and scales with the system size, as expected for strongly localized  states. Unlike the typical case of Anderson disorder~\cite{schleede_comment_2010},  $R(E)\rightarrow 1$ close to the van Hove singularity, indicating that  localization is restricted to the vicinity of the Dirac point. 

\begin{figure}[bp]
     \centering\includegraphics[width=1\linewidth,clip]{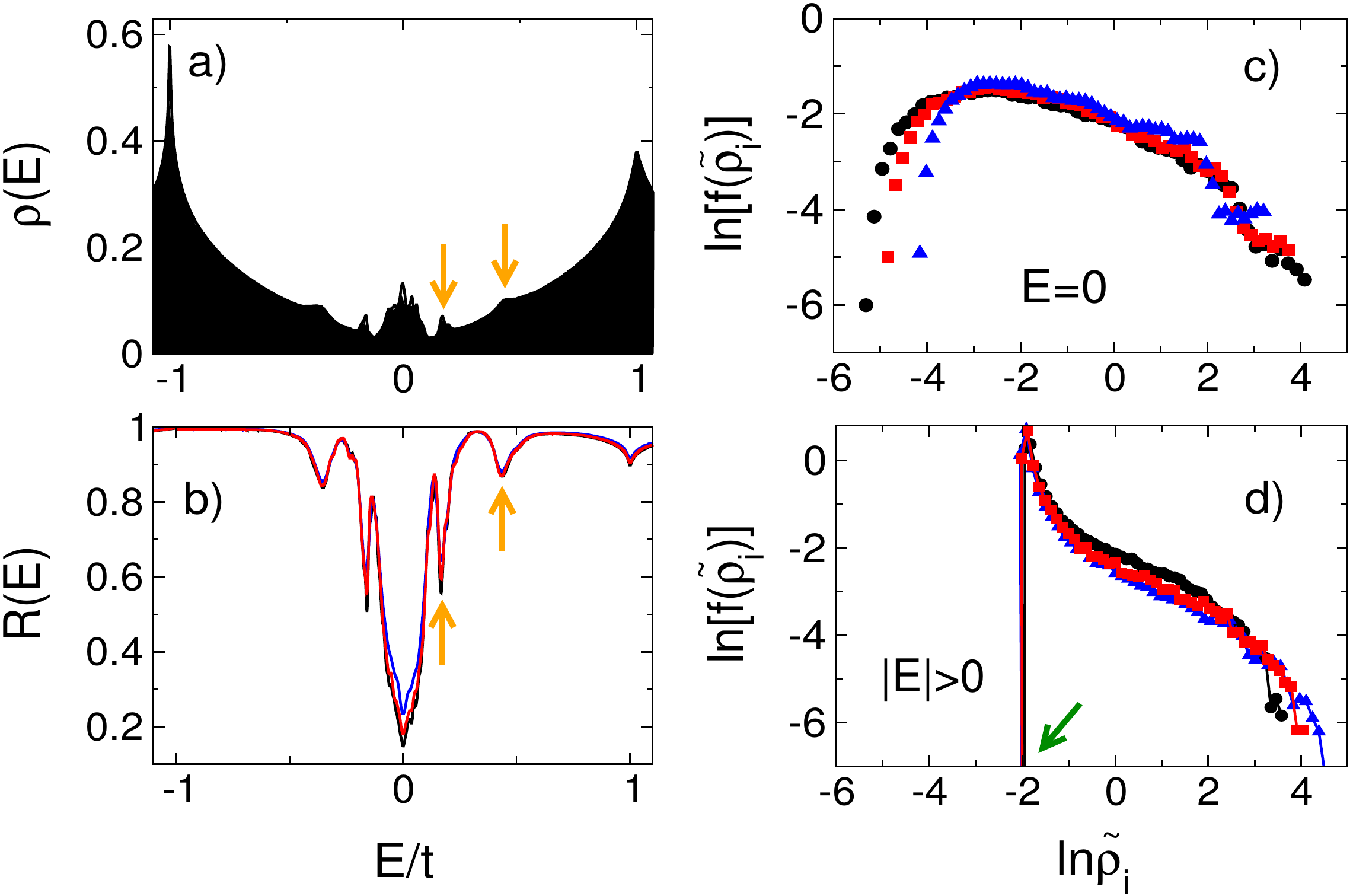}
     \caption{ (a) DOS for resonant $s$-wave orbitals. (b) $R(E)$ for different system sizes ($p=0.05$). (c) Normalized LDOS distribution function at the Dirac point ($E=0$) and (d) away from it ($E=0.1t$).  The system sizes in panels b-d are $D=2\times L\times L$ with $L=300$ (blue triangles), 600 (red squares) and 1200 (black dots) }
     \label{fig:sdos}
\end{figure}

This analysis is consistent with the distribution functions of  $\ln\tilde{\rho}_i$ for $E=0.1 t$, shown in Fig. ~\ref{fig:dxydos}c, which  are described by  Gaussian distributions \cite{schubert_distribution_2010}. In log-scale, the curves have a parabolic shape, which is expected for AL. Moreover, these curves  also scale with the system size.  In agreement with the AL  scenario, the peak of the log-normal curve is shifted towards lower densities when the system size is increased, indicating an insulating state in the thermodynamic limit ~\cite{schubert_distribution_2010}. The finite size scaling analysis, depicted in  Fig.~\ref{fig:dxydos}d, shows that the tails of the three curves collapse into a single universal curve.  The same strong localization features were observed for all energies in the range of the plateau in $R(E)$ (Fig.\ref{fig:dxydos}b). This analysis remains valid  for different values of   $\tau \in [-0.67t,-0.4t]$,  around the resonant condition.  Both $R$ and the distribution function vary very little with $p$ at the Dirac point, even for  concentrations as small as $p=0.001$. The localization nevertheless seems to disappear for states above and below the Dirac point.  The localization phenomenon is  robust not only as a function of the impurity concentration but also in the presence of a random admixture of resonant adatoms with different $d$-wave orbital symmetries.

The physics of AL changes dramatically for adatoms in the symmetric class, where all hopping amplitudes in the plaquette are the same, $\tau_{ij}=\pm\tau$.  Fig. \ref{fig:sdos}a  shows a resonance peak in the DOS  at the Dirac point for $\tau=0.425t$. This peak is accompanied by two symmetric satellite peaks, indicated by the arrows.  In Fig. \ref{fig:sdos}b, $R(E)$ has a pronounced minimum at the Dirac point together with two additional minima at the energies of the extra peaks in the DOS. In contrast with the asymmetric case, $R(E)$ shows a significant variation with the system size \emph{only} at the Dirac point and is size independent at nearly all other energies. At the Dirac point, the distribution functions shown in Fig. \ref{fig:sdos}c  clearly scale with the size of the system. Although they are not parabolic, they resemble the distribution functions of  Fig.~\ref{fig:dxydos}c, and have substantially more weight at sites with very low densities, indicating that the system is strongly localized. Those AL features remain robust for $\tau\in [0.39,0.46]t$, in the vicinity of the resonant condition. 
\begin{figure}[tbp]
\centering
         \includegraphics[width=\linewidth,clip]{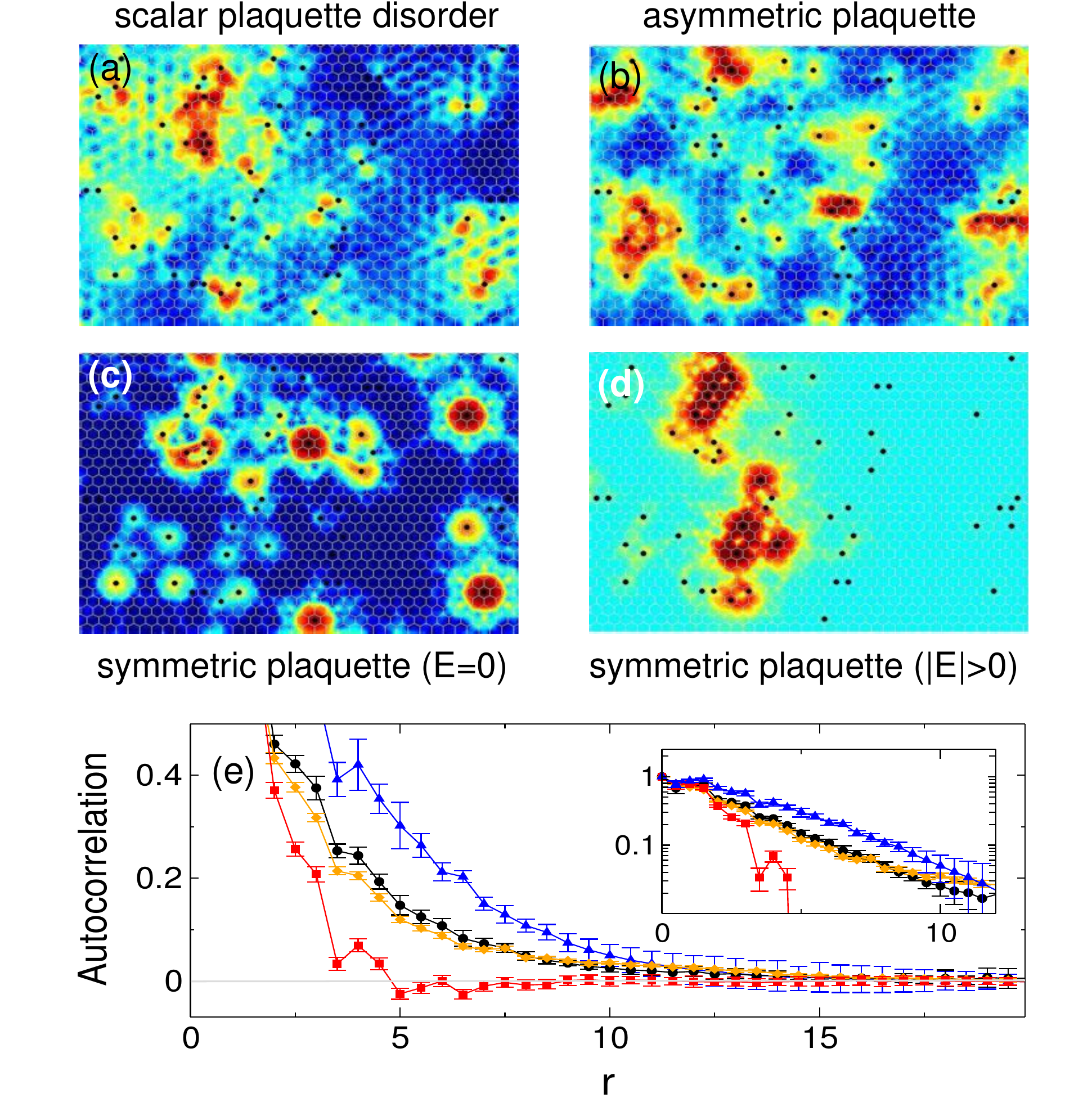}
      \caption{Normalized LDOS at the Dirac point  for: a) scalar plaquete disorder and b) asymmetric plaquette disorder  for $d$ orbitals.  c) symmetric plaquette disorder ($s$-wave orbital), at the Dirac point and  d) away from it ($E=0.1t$, see text). e) Autocorrelation functions for  the LDOS of panels b-d vs. distance $r$ in lattice parameters. Inset: log scale. Blue triangles: $s$-wave, $E=0.1t$; red squares: $s$-wave at $E=0$. Black dots: $d_{xy}$ at $E=0$ and (orange diamonds) $E=0.1t$.   }

     \label{fig:sldos}
\end{figure}

Away from the Dirac point,  the system crosses over to an exotic metallic state. The distribution functions shown in Fig. \ref{fig:sdos}d  have a sharp lower bound  at $\ln \tilde{\rho}_i \sim -2$ (green arrow), which does \emph{not} scale with the system size, and hence indicates metallic behavior in the thermodynamic limit.   At the same time, they show a  power-law tail for large values of $\tilde{\rho}_i$, which is characteristic of a strongly inhomogeneous system. In Mott insulators, those features have been linked to a metallic state that is a precursor to electronic Griffiths phases \cite{andrade}. This exotic metallic state survives  in the presence of scalar plaquette disorder  for $|\delta\mu| < \delta\mu_c \sim 1.4 t$. For $|\delta\mu| > \delta\mu_c$, a typical Anderson transition driven by  $\delta\mu$ appears in the vicinity of the Dirac point. Although large, this critical scalar plaquette potential is much smaller than the  {\it on-site} potential required to localize charge carriers in graphene, which is of the order of $15t$~\cite{pereira2}. 

In Fig. \ref{fig:sldos} we  compare  LDOS patterns in real space for different kinds of plaquette disorder. Panel \ref{fig:sldos}a displays the LDOS  for strong scalar disorder at $E=0$, for  $\delta \mu=1.6 t$ and $\tau=0$. The pattern  is similar to the asymmetric plaquette disordered case at $\delta\mu=0$ and $\tau=-0.56t$, shown in Fig. \ref{fig:sldos}b  for the case of $d$-wave orbitals. 
In the symmetric case, at   $E=0$, $\delta \mu=0$ and $\tau=0.425t$, the LDOS  has a very different structure, and shows characteristic puddles with the radius of $\sim2.3a$, with $a$ the lattice parameter, around isolated adatoms (see panel 4c).  Away for the Dirac point (panel \ref{fig:sldos}d), isolated adatoms nearly decouple from the bath and large puddles appear around \emph{rare} clusters of adatoms, leading to a metallic state.

In panel \ref{fig:sldos}e, we show the autocorrelation function 
\begin{equation}
C(\mathbf{r},E)=1/(2\pi D) \sum_{i} \delta \rho(\mathbf{R}_i,E) \delta \rho(\mathbf{R}_i+\mathbf{r},E)
\end{equation}
for panels b-d,  where $\delta\rho(\mathbf{R}_i) \equiv \rho(\mathbf{R}_i)- \rho$ is the variation of the LDOS away from the average~\cite{STMscience}.  The curves corresponding to the asymmetric plaquette case ($d_{xy}$)  for  $E=0$ and $0.1t$ decay exponentially, as indicated in the inset. The two curves have the same correlation length $\xi\sim 2.6a$, and are indistinguishable (black dots and orange diamonds). For the symmetric case, the metallic state ($E=0.1t$) shows a slower exponential decay (blue triangles) with $\xi\sim 5.5a$, crossing over to a localized state at $E=0$ (red squares). The autocorrelation function of the latter decays much faster than in conventional AL, with rapid oscillations around $\delta\rho=0$, suggesting a strongly localized state.

\section{Experimental observation and conclusions}

In summary,  we proposed a plaquette disorder potential, that describes the local hopping processes mediated by an adatom that sits at the center of the honeycomb plaquette. We showed that the problem of AL for resonant adatoms in graphene can be separated in two symmetry classes based on whether the resonant orbitals of the impurity break or preserve the sublattice point group symmetry. While the first class of localization (asymmetric) is more conventional, the second one (symmetric class) is anomalous. In the latter, we show that destructive interference effects among different hopping paths in the plaquette of the adatom produce a singular localized state at the Dirac point, and away from it, an exotic metallic state with rare charge droplets that has the same signatures that were previously identified in a precursor to an electronic Griffiths phase.

The experimental characterization of AL can be done through STS probes \cite{STMscience}, which can scan the LDOS in the vicinity of the Dirac point. Localization features can also be observed in transport measurements through the scaling of the conductance with the system size \cite{abrahams_scaling_1979}. Recent \textit{ab initio} calculations indicate that the 4$s$ orbital of Cu adatoms forms a midgap state, while the 3$d$ orbitals of  Co, Fe and V adatoms may  display resonances near the Dirac point \cite{Cao, bonding}. In addition, STS measurements reported that Ni adatoms form a midgap state with $s$-wave orbital symmetry \cite{Gyamfi, Wehling13}. Those results suggest that  Ni and Cu adatoms are good candidates for the observation of the Anderson transition in the symmetric class, which leads to a strongly localized state at the Dirac point only, while  Co, Fe and V adatoms may lead to AL in the asymmetric class, where localization is expected over a finite window of energy around the Dirac point.  In all cases,  excluding Ni, the orbitals in the resonant levels are spin polarized  \cite{Cao, bonding}. Above the Kondo temperature, the exchange coupling between the itinerant and local spins can further enhance AL effects.

We acknowledge F. Guinea, K. Mullen, A. H. Castro Neto and E. Mucciolo for discussions. B.U. acknowledges University of Oklahoma for financial support.  T.G.R and J.H.G acknowledge Brazilian agencies CNPq, FAPERJ and "INCT de nanoestruturas de carbono" for financial support.

\appendix

\section{Resonant condition \label{sec-App}}

In momentum space representation, the Hamiltonian of graphene plus
one single impurity can be written as 
\begin{equation}
\mathbf{H}=\left(\begin{array}{cc}
\mathbf{H}_{g} & \mathbf{V}\\
\mathbf{V}^{\dagger} & \mathbf{H}_{f}
\end{array}\right),\label{eq:H1}
\end{equation}
where $\mathbf{H}_{g}$ is a $2\times2$ matrix in the graphene sublattice
basis, $\Psi=(\psi_{a},\psi_{b})$. The second block matrix, $\mathbf{H}_{f}=\epsilon_{0} \delta_{m,m^\prime}$, 
is a diagonal matrix in the $f_m$ basis of localized electrons. 
\begin{equation}
\mathbf{V}_{\mathbf{p}}^{(m)}=\left(\begin{array}{c}
V_{a,\mathbf{p}}^{(m)}\\
V_{b,\mathbf{p}}^{(m)}
\end{array}\right)\label{eq:V}
\end{equation}
is the $2\times1$ hybridization matrix for a given orbital (indexed by $m$), and matrix elements 
\begin{eqnarray}
V_{a,\mathbf{p}}^{(m)} & = & \sum_{j\in I_{A}}V_{j}^{(m)}\mbox{e}^{-i\mathbf{p}\cdot\mathbf{R}_{j}}\label{eq:va}\\
V_{b,\mathbf{p}}^{(m)} & = & \sum_{j\in I_{B}}V_{j}^{(m)}\mbox{e}^{-i\mathbf{p}\cdot\mathbf{R}_{j}}\label{eq:vb}
\end{eqnarray}
where $\mathbf{p}$ is the momentum, $j\in I_{A}$ describes the hybridization
amplitudes of the impurity with the three nearest carbon atoms on sublattice $A$, $(V_{j}=V_{1,3,5})$,
$j\in I_{B}$ the hybridization amplitudes with the other three nearest carbon atoms in sublattice $B$ $(V_{j}=V_{2,4,6})$,
and $\mathbf{R}_{j}$ describes the position of the six carbon atoms
in the plaquete centered at the origin. 

The \emph{exact} Green's function of graphene in the presence of a
single impurity is \cite{bruno2011}: 
\begin{widetext}
\begin{equation}
\mathbf{G}(\mathbf{p},\mathbf{p}^{\prime},\omega)=\delta_{\mathbf{p},\mathbf{p}^{\prime}}\mathbf{G}^{0}(p)+\sum_{m,m^\prime}\left[\mathbf{G}^{0}(p)\cdot\mathbf{V}_{\mathbf{p}}^{(m)}\cdot G_{f,mm^\prime}(\omega)\cdot\mathbf{V}_{\mathbf{p}^{\prime}}^{\dagger(m^\prime)}\cdot\mathbf{G}^{0}(p^{\prime})\right]\label{eq:Gpp-1}
\end{equation}
\end{widetext}
where the sum over $m$ includes \emph{all} degenerated orbitals with
energy $\epsilon_{0}$, 
\begin{equation}
\mathbf{G}^{0}(\mathbf{p},\omega)=[\omega-\mathbf{H}_{g}(\mathbf{p})+i0^{+}]^{-1}\label{eq:G0}
\end{equation}
is the $2\times2$ matrix of the unperturded Green's function, and
$G_{f,mm^\prime}(\tau)=-\langle T[f_m(\tau)f_{m^\prime}^{\dagger}(0)\rangle$ is the
retarded Green's function of the localized electrons, 
\begin{equation}
G_{f,mm^\prime}(\omega)=\left[(\omega-\epsilon_{0})\delta_{m,m^\prime}-\Sigma_{f,mm^\prime}(\omega)+i0^{+}\right]^{-1}\label{Gff2}
\end{equation}
with
\begin{equation}
\Sigma_{f,mm^\prime}(\omega)=\sum_{\mathbf{p}}\mathbf{V}_{\mathbf{p}}^{\dagger(m)}\cdot\mathbf{G}^{0}(\mathbf{p},\omega)\cdot\mathbf{V}_{\mathbf{p}}^{(m^\prime)}\label{SEH0}
\end{equation}
the corresponding self-energy, which is a matrix in the degenerate space of the orbitals.

From Eq. (\ref{eq:Gpp-1}), the resonance in the LDOS nearby the impurity at the Dirac point,
\begin{equation}
\rho(0)=-\frac{1}{\pi}\mathrm{tr}\sum_{\mathbf{p},\mathbf{p}^{\prime}}\mbox{Im}\left[\mathbf{G}(\mathbf{p},\mathbf{p}^{\prime},0)\right],\label{eq:rho}
\end{equation}
follows from a pole in the denominator of $G_{f}(0)$. In the non-degenerate case, where the orbital indexes $m,m^\prime$ can be dropped, $-\epsilon_{0}-\Sigma_{f}(0)+i0^{+}=0$.

The resonant condition for a single impurity corresponds to the effective
hopping parameter 
\begin{equation}
\tau_0=-\frac{V^{2}}{\epsilon_{0}}= V^2 \left[\mbox{Re}\Sigma_{f}(0)\right]^{-1},\label{eq:tau}
\end{equation}
where the imaginary part of $\Sigma_{f}$ accounts for the level broadening
$\Delta(0)=\mbox{Im}\Sigma_{f}(0)$, which sets the width of the resonance.
In graphene, 
\begin{equation}
\mathbf{H}_{g}(\mathbf{p})=-\left(\begin{array}{cc}
0 & t \phi_{\mathbf{p}}\\
t \phi^*_{\mathbf{p}}& 0
\end{array}\right),
\end{equation}
where $\phi_{\mathbf{p}} = \sum_{j\in I_A} \exp{i \mathbf{p} \cdot \mathbf{R}_j }$.  From Eq. (\ref{SEH0}), the self-energy for a single non-degenerate orbital is 
\begin{equation}
\mbox{Re}\Sigma_{f}^{(m)}(0)=\sum_{\mathbf{p}} \mathbf{V}^{(m)}\cdot\mathbf{G}^{0}(\mathbf{p},0)\cdot\mathbf{V}^{(m)}.
\end{equation}

For a symmetric orbital ($s$-wave), where $V_{i\in I}=V$, $(m=0)$, 
\begin{equation}
\mbox{Re}\Sigma_{f}^{(0)}(0)\approx2.346\, V^2/t,
\end{equation}
This value is very close to the resonant hopping that produces a midgap state,  $\tau_0=0.425t$, which was computed numerically in the main text from the effective graphene-only Hamiltonian.
 For a non-degenerate $d_{xy}$-wave orbital, where
$V_{1,4}=0$ and $V_{2,5}=-V_{3,6}=\sqrt{3}V/2$, $(m=2)$, 
or equivalently for a $d_{x^2-y^2}$ orbital, where $V_{1,4}=V$, and  $V_{2,3,5,6}=-V/2$ ($m=-2$), 
\begin{equation}
\mbox{Re}\Sigma_{f}^{(\pm2)}(0)\approx-1.782 \,V^2/t,
\end{equation}
which gives $\tau_0\approx-0.561t$. Those results also agree with the values extracted numerically from the effective graphene-only Hamiltonian for $d_{xy}$ and $d_{x^2-y^2}$ orbitals, $\tau= -0.56t$.

In the degenerate case, the condition for a mid gap state is 
$\mathrm{Det}[-\epsilon_0 - \bm{\Sigma}_{f}(0) +i0^+]=0. $
For a superposition of two degenerate $d$-wave
orbitals, $(m=\pm2)$, where $\bm{\Sigma}_f$ is a $2\times2$ matrix, 
one recovers the non-degenerate $d$-wave result.

\section{Graphene-only Hamiltonian \label{sec-go}}

The \emph{exact} Green's function of graphene in the presence of a
single impurity can be written as
\begin{equation}
\mathbf{G}=\left[i\omega-\mathbf{H}_{eff}\right]^{-1}=[\mathbf{G}_{0}^{-1}-\bm{\Sigma}]^{-1}\label{eq:G2}
\end{equation}
where $\bm{\Sigma}$ is the self-energy. Using the identity: $$1+\mathbf{A}=[1-\mathbf{A}\cdot(1+\mathbf{A})^{-1}]^{-1},$$
one can extract the self energy by combining Eq. (\ref{eq:Gpp-1})
and Eq. (\ref{eq:G2}), 
\begin{equation}
\bm{\Sigma}(\mathbf{p},\mathbf{p^{\prime}},\omega)= 
\sum_{\mathbf{p}^{\prime\prime} } \mathbf{G}_{0}^{-1}(p)\cdot\bm{\Gamma}_{\mathbf{p},\mathbf{p}^{\prime\prime}}(\omega)\cdot \bm{\Lambda}_{\mathbf{p}^{\prime},\mathbf{p}^{\prime\prime}}(\omega)\cdot\mathbf{G}_{0}^{-1}(p^{\prime}),\label{eq:Sigma2} 
\end{equation}
where 
\begin{equation}
\bm{\Gamma}_{\mathbf{p},\mathbf{p}^{\prime}}(\omega)=\sum_{m,m^\prime}\left[\mathbf{G}^{0}(p)\cdot\mathbf{V}_{\mathbf{p}}^{(m)}\cdot G_{f,mm^\prime}(\omega)\cdot\mathbf{V}_{\mathbf{p}^{\prime}}^{\dagger(m^\prime)}\cdot\mathbf{G}^{0}(p^{\prime})\right]\label{eq:Gamma}
\end{equation}
 and 
 \begin{equation}
 \bm{\Lambda}_{\mathbf{p}^{\prime},\mathbf{p}^{\prime\prime}}(\omega)=[\delta_{\mathbf{p}^{\prime\prime}\mathbf{p}^{\prime}}+\mathbf{G}_{0}^{-1}(p^{\prime\prime})\cdot\bm{\Gamma}_{\mathbf{p}^{\prime\prime}\mathbf{p}^{\prime}}(\omega)]^{-1}.
 \end{equation}

At zero energy, $(\omega=0)$, where ${\rm Im}\mathbf{\bm{\Sigma}}(0)=0$,
the effective graphene-only Hamiltonian is 
\[
\mathbf{H}_{eff}(\mathbf{p},\mathbf{p}^{\prime})=\delta_{\mathbf{p,}\mathbf{p}^{\prime}}\mathbf{H}_{0}(\mathbf{p})+\bm{\Sigma}(\mathbf{p},\mathbf{p}^{\prime},0).
\]
In explicit form, 
\begin{widetext}
\begin{equation}
\mathbf{H}_{eff}(\mathbf{p},\mathbf{p}^{\prime})=\delta_{\mathbf{p,}\mathbf{p}^{\prime}}\mathbf{H}_{0}(\mathbf{p})+\sum_{\mathbf{p}^{\prime\prime}}\mathbf{H}_{0}(\mathbf{p})\cdot\bm{\Gamma}_{\mathbf{p},\mathbf{p}^{\prime\prime}}\cdot[\delta_{\mathbf{p}^{\prime\prime}\mathbf{p}^{\prime}}+\mathbf{H}_{0}(\mathbf{p}^{\prime\prime})\cdot\bm{\Gamma}_{\mathbf{p}^{\prime\prime},\mathbf{p}^{\prime}}]^{-1}\cdot\mathbf{H}_{0}(\mathbf{p}^{\prime}).\label{eq:Heff4}
\end{equation}
\end{widetext}
In leading order in perturbation in the hybridization $V$, 
\begin{equation}
\mathbf{H}_{eff}(\mathbf{p},\mathbf{p}^{\prime})=\delta_{\mathbf{p},\mathbf{p}^{\prime}}\mathbf{H}_{0}(\mathbf{p})-\frac{1}{\epsilon_{0}}\sum_{m}\mathbf{V}_{\mathbf{p}}^{(m)}\cdot\mathbf{V}_{\mathbf{p}^{\prime}}^{\dagger(m)}+\mathcal{O}(V^{3})\label{eq:Heff3}
\end{equation}
Equivalently, taking the Fourier transform to real space, the effective graphene-only Hamiltonian is
\begin{equation}
\mathcal{H}_{eff}=\mathcal{H}_{0}+\sum_{(i,j)\in I}\tau_{i,j}\,\psi^{\dagger}(\mathbf{R}_{i})\psi(\mathbf{R}_{j}),\label{eq:HeffV2}
\end{equation}
where $(i,j)\in I$ index the six carbon atoms in the plaquete around
the impurity, and 
\begin{equation}
\tau_{i,j}=-\frac{1}{\epsilon_{0}}\sum_{m}V_{i}^{(m)}V_{j}^{*(m)}\label{eq:tau2}
\end{equation}
is the effective hopping mediated by the impurity.

\end{document}